\documentstyle[12pt,epsfig]{article}

\newlength{\dinwidth}
\newlength{\dinmargin}
\def\docnum#1{\hbox to \hsize{\hskip123mm\hbox{#1}\hss}}
\def\date#1{\edef\@temp{#1}\ifx\@temp\@empty\def\@temp{\today}\fi
\hbox to \hsize{\hskip123mm\hbox{\@temp}\hss}}
\def\title#1{\vskip 0.8in plus 2in\begin{center}%
{\Large\bf#1\par}\vskip1.5em\end{center}\vskip 1in}
\def\@makefnmark{\hbox{$^{\@thefnmark)}$}}
\def\author#1{
\setcounter{footnote}{0}\def\@currentlabel{}%
\begingroup\def\thefootnote{\arabic{footnote}}
\def\@makefnmark{\hbox{$^{\@thefnmark)}$}}
\global\@topnum\z@ \large\begin{center}{\lineskip.5em
\begin{tabular}[t]{c}#1\end{tabular}\par}
\end{center}\par\vskip1.5em\@thanks\endgroup}

\def\abstract{\vskip0.8in plus 
3in\begin{center}{\large\bf Abstract}\end{center}\quotation}

\newcommand{\gs}   {$\gamma_{\rm s}\ $}
\newcommand{\ppb}  {$\rm{p\bar{p}}\;$}
\newcommand{\ssb}  {$\rm{s\bar{s}}\;$}
\newcommand{\pp}   {$\rm{pp}$ }
\newcommand{\ls}   {$\lambda_{\rm s}\ $}
\newcommand{\ee}   {${\rm e}^+{\rm e}^-$}
\newcommand{\qi}   {{\mathbf{q}}_i}
\newcommand{\vexp} {$VT^3\exp{(-0.7 {\rm GeV}/T)}$}
\newcommand{\muv}  {{\mbox{\boldmath ${\mu}\!$ \unboldmath} }}

\begin{document}

\begin{titlepage}
\flushright{BI-TP 00/02}
\vspace{-0.35cm}
\flushright{DFF 349/02/2000}
\vspace{-1.6cm}
\title{Features of particle multiplicities and strangeness production
in central heavy ion collisions between 1.7$A$ and 158$A$ GeV$/c$.}
\vspace{-2.6cm}
\begin{center}
\large{F. Becattini}\\
{\it Universit\`a di Firenze and INFN Sezione di Firenze,\\
Largo E. Fermi 2, I-50125, Florence, Italy}\\
\large{J. Cleymans}\\
{\it Department of Physics, University of Cape Town,\\
Rondebosch 7701, Cape Town, South Africa}\\
\large{A. Ker\"anen, E. Suhonen}\\
{\it Department of Physical Sciences, University of Oulu,\\
FIN-90571 Oulu, Finland}\\
\large{K. Redlich}\\
{\it Institute of Theoretical Physics, University of Wroclaw,\\
PL-50204 Wroclaw, Poland.}\\
\end{center}
\vspace{-1.5cm}
\begin{abstract}

A systematic study is performed of fully integrated particle multiplicities
in central Au--Au and Pb--Pb collisions at beam momenta of 1.7$A$ GeV/c, 11.6$A$ 
GeV/c (Au--Au) and 158$A$ GeV/c (Pb--Pb) by using a statistical-thermal model.
The close similarity of the colliding systems makes it possible to study heavy 
ion collisions under definite initial conditions over a range of centre-of-mass
energies covering more than one order of magnitude.
In order to further study the behaviour of strangeness production, an updated study
of Si--Au collisions at 14.6$A$ GeV is also presented. The data analysis has
been performed with two completely independent numerical algorithms giving closely
consistent results. We conclude that a thermal model description of particle
multiplicities, with additional strangeness suppression, is possible for each 
energy.
The degree of chemical equilibrium of strange particles and the relative
production of strange quarks with respect to u and d quarks are higher than in \ee,
\pp and \ppb collisions at comparable and even at lower energies.
The behaviour of strangeness production as a function of centre-of-mass energy and
colliding system is presented and discussed. The average energy per hadron in the
comoving frame is close to 1 GeV per hadron despite the fact that the energy
increases more than 10-fold.
\end{abstract}
\centerline{PACS:24.10.Pa,25.75.Dw,25.75.-q}
\end{titlepage}

\section{Introduction}

After scouring results from relativistic heavy ion collisions at many different
energies over several years \cite{heinz} some common traits are starting to 
emerge. Indeed, statistical-thermal models have proved to be able to reproduce 
relative particle multiplicities in a satisfactory manner by using two or three 
relevant parameters: temperature, baryon chemical potential and a possible 
strange-quark suppression parameter, \gs \cite{gs}. Such an analysis has been 
performed by many authors for heavy ion collisions data from CERN SPS, from 
Brookhaven AGS and also from GSI SIS. In this paper we present a simultaneous 
analysis of data from several different collisions, with emphasis on the 
similarity of the colliding system in order to study the behaviour of parameters 
as a function of centre-of-mass energy within one framework. Hence, we have 
focussed our attention on central Au--Au collisions at beam momenta of 1.7$A$ 
GeV/c (SIS) \cite{oeschler}, 11.6$A$ GeV/c (AGS) \cite{ogilvie} and on central 
Pb--Pb collisions at 158$A$ GeV/c (SPS) \cite{stock}.
As far as the choice of data (and, consequently, colliding system) is concerned,
our leading rule is the availability of full phase space integrated multiplicity
measurements because a pure statistical-thermal model analysis of particle yields,
without any consideration of dynamical effects, {\em may} apply only in this case
\cite{prc}. Such data, however, exist only in a few cases and whenever legitimate
we have extrapolated spectra measured in a limited rapidity window to full phase
space. The use of extrapolations is more correct than using data over limited
intervals of rapidity, especially in the framework of a purely statistical-thermal
analysis without a dynamical model. Moreover, the usual requirement of zero 
strangeness ($S=0$) demands fully integrated multiplicities because strangeness 
does not need to vanish in a limited region of phase space.

A point of considerable interest in heavy ion collisions is the enhanced
production of strange quarks per u, d quark with respect to elementary collisions 
\cite{bgs} like \ee, \pp, \ppb. This could be related to properties of the system 
at the parton level prior to hadronisation \cite{heinz,stock,bgs,stockplb}. In 
order to further study strangeness production and enhancement at low energy, we 
also present a new analysis of Si--Au collisions at 14.6$A$ GeV (AGS) using only 
multiplicities obtained from fully integrated phase space distributions.
This also allows to cross-check results of previous analyses \cite{BNL-PBM,BNL-thews,heppe}
performed using limited rapidity interval data. In particular, we have included
the $4\pi$ pion multiplicity \cite{abbottnew} and results presented in \cite{Si2}.
In order to assess the consistency of the results obtained, we have performed the
statistical-thermal model analysis by using two completely independent numerical
algorithms whose outcomes turned out to be in close agreement throughout.
 
Similar analyses have been recently made by other authors (see e.g. \cite{heppe,letessier});
however, both the model and the used data set differ in several important details,
such as the assumption of full or partial equilibrium for some quark flavours,
the number of included resonances, the treatment of resonance widths, inclusion
or not of excluded volume corrections, treatment of flow, corrections due to limited
rapidity windows etc. Because of these differences it is difficult to trace the 
origin of discrepancies between different results. We hope that the present analysis, 
covering a wide range of beam energies using a consistent treatment, will make it 
easier to appreciate the energy dependence of the various parameters such as temperature 
and chemical potential.

\section{Data set and model description}

As emphasized in the introduction, in the present analysis we use the most recent
available data, concentrating on fully integrated particle yields and discarding
data that have been obtained in limited kinematic windows. The only exceptions to
this rule are the $\overline\Lambda/\Lambda$ and $\overline{\rm p}/{\rm K}^-$ ratios
in Si--Au collisions \cite{Si2,Si3} which were not available in full phase space.
It has been decided to keep them as they are the only available recent measurements
involving antibaryons.

We have derived integrated multiplicities of $\mathrm{\pi}^+$, $\Lambda$ and proton
in Au--Au collisions at AGS by extrapolating published rapidity distributions
\cite{piAu,LAu,pAu} with constrained mid-rapidity value ($y_{\rm NN}$=1.6). For
proton and $\Lambda$ we have fitted the data to Gaussian distributions, whilst for
$\pi^+$ we have used a symmetric flat distribution at midrapidity with Gaussian-shaped
wings on each side; the point at which the Gaussian wing and the plateau connect
is a free parameter of the fit. The fits yielded very good $\chi^2$'s/dof: 0.27, 1.24
and 1.00 for $\pi^+$, proton and $\Lambda$ respectively. The integrated multiplicities
have been taken as the area under the fitted distribution between the minimal 
$y_{\rm min}$ and maximal $y_{\rm max}$ values of rapidities for the reactions
${\rm N N} \rightarrow \pi {\rm N N}$, ${\rm N N} \rightarrow \Lambda {\rm K}$ for
pions and $\Lambda$'s respectively; the difference between these areas and the total
area has been taken as an additional systematic error. The area between $y_{\rm min}$
and $y_{\rm max}$ amounts to practically 100\% of the total area for pions and about
95\% for $\Lambda$'s. Ref.~\cite{LAu} quotes an additional experimental systematic
error of 10\% on $\Lambda$ multiplicity that we have added in quadrature. Hence we
obtain:

\begin{eqnarray}
  \langle \pi^+ \rangle   &= &133.7 \pm 9.93 \nonumber \\
  \langle \Lambda \rangle &= &20.34 \pm 1.36 \pm 1.23 \pm 2.03
\label{eqn:fit}
\end{eqnarray}
where the first error is the fit error, the second is the systematic error
due to the variation of integration region and the third is the experimental
systematic error. As to protons, the extracted rapidity interval corresponding to
the reaction N N $\rightarrow$ N N is only 79\% of the total Gaussian area. The
difference between the two areas is too large to be considered as an additional
error; thus, in order to reduce the uncertainty, we have decided to take the ratio
${\rm p}/\pi^+$ extracted in the above rapidity interval rather than the proton
multiplicity itself. This yields:

\begin{equation}
  \langle {\rm p}/\pi^+ \rangle = 1.234 \pm 0.126  
\label{eqn:fit2}
\end{equation}
where the error includes both the fit error and an error stemming from a 10\%
systematic uncertainty quoted in ref.~\cite{pAu}.

We have not included data on deuteron production because of the possible
inclusion of fragments in the measured yields. This is particularly dangerous at 
low (SIS) energies where inclusion or not of deuterons modifies thermodynamic 
quantities like $\epsilon /n$ \cite{prl}.

The data analysis has been performed within an ideal hadron gas grand-canonical
framework for Pb--Pb at SPS and Au--Au at AGS whereas for Au-Au at SIS and Si--Au 
at AGS we have required the exact conservation of strangeness instead of using a 
strangeness chemical potential (see the discussion later in the text); in both 
cases we have used a supplementary strange quark fugacity $\gamma_{\rm s}$. In 
the grand-canonical approach, the overall average multiplicities of hadrons and 
hadronic resonances are determined by an integral over a statistical distribution:      

\begin{equation}
\langle n_i \rangle = (2 J_i + 1)\frac{V}{(2\pi)^3} \int {\rm d}^3p \,\,
\frac{1}{\gamma_{\rm s}^{-s_i} \exp{[(E_i-\muv \cdot \qi)/T]} \pm 1}
\label{eqn:thermaldist}
\end{equation}
where $\qi$ is a three-dimensional vector with electric charge, baryon number and
strangeness of hadron $i$ as components; \muv the vector of relevant chemical
potentials; $J_i$ the spin of hadron $i$ and $s_i$ the number of valence strange
quarks in it; the $+$ sign in the denominator is relevant for fermions, the $-$
for bosons. This formula holds in case of many different statistical-thermal
systems (i.e. clusters or fireballs) having common temperature and \gs but 
different arbitrary momenta, provided that the probability of realizing a given 
distribution of quantum numbers among them follows a statistical rule \cite{bgs,beca}. 
In this case $V$ must be understood as the sum of all cluster volumes measured in 
their own rest frame. Furthermore, since both volume and participant nucleons may 
fluctuate on an event by event basis, $V$ and \muv (and maybe $T$) in 
Eq.~(\ref{eqn:thermaldist}) should be considered as average quantities \cite{bgs}.

The overall abundance of a hadron of type $i$ to be compared with experimental
data is determined by the sum of Eq.~(\ref{eqn:thermaldist}) and the contribution
from decays of heavier hadrons and resonances:

\begin{equation}
  n_i = n_i^{\rm primary} + \sum_j {\rm Br}(j\rightarrow i) n_j
\label{eqn:sum}
\end{equation}
where the branching ratios Br$(j\rightarrow i)$ have been taken from the 1998
issue of the Particle Data Table \cite{PDG}.

It must be stressed that the unstable hadrons contributing to the sum in
Eq.~(\ref{eqn:sum}) may differ according to the particular experimental definition.
This is a major point in the analysis procedure because quoted experimental
multiplicities may or may not include contributions from weak decays of hyperons 
and K$^0_S$. We have included all weak decay products in our computed
multiplicities except in Pb--Pb collisions on the basis of relevant statements
in ref.~\cite{sikler} and about antiproton production in refs.~\cite{Si3,antipAu}.
It must be noted that switching this assumption in Au--Au at SIS and AGS
does not affect significantly the resulting fit parameters whereas it does in
Si--Au.

The overall multiplicities of hadrons depend on several unknown parameters
(see Eq.~(\ref{eqn:thermaldist})) which are determined by a fit to the data.
The free parameters in the fit are $T$, $V$, \gs and $\mu_B$ (the baryon chemical 
potential) whereas $\mu_S$ and $\mu_Q$, i.e. the strangeness and electric
chemical potentials, are determined by using the constraint of overall vanishing 
strangeness and forcing the ratio between net electric charge and net baryon number 
$Q/B$ to be equal to the ratio between participant protons and nucleons. The latter 
is assumed to be $Z/A$ of the colliding nucleus in Au--Au and Pb--Pb while it 
has been calculated to be 0.43 for central Si--Au collisions by means of a 
geometrical model.

As we have mentioned before, for SIS Au-Au and AGS Si--Au data we have required 
the exact conservation of strangeness instead of using a strangeness chemical 
potential. This gives rise to slightly more complex calculations \cite{canstran} 
which are necessary owing to either very small strange particle production (Au--Au) 
or a relatively small system size (Si--Au). The difference between the 
strangeness-canonical and pure grand-canonical calculationis of multiplicities of K 
and $\Lambda$ for the final set of thermal parameters (see Table~1) turns out to 
be around 2-3\% for K and $\Lambda$ in Si--Au but it is as large as a factor 15 
in Au--Au at 1.7$A$ GeV/c.

Owing to few available data points in SIS Au--Au collisions, we have not
fitted the volume $V$ nor the \gs  therein. The volume has been assumed to be
$4\pi r^3/3$ where $r = 7$ fm (approximately the radius of a Au nucleus) while
\gs has been set to 1, the expected value for a completely equilibrated hadron
gas. Since we have performed a strangeness-canonical calculation here, the yield
ratios involving strange particles are not independent of the chosen volume value
as in the grand-canonical framework. Thus, in this particular case, $V$ is
meant to be the volume within which strangeness is conserved (i.e. vanishing)
and not the global volume defining overall particle multiplicities as in
Eq.~(\ref{eqn:thermaldist}). Also, in order to test the dependence of this assumption 
on our results, we have repeated the fit by varying $V$ by a factor 2 and 0.5 in 
turn.
         
The yields of resonances have been calculated by integrating Eq.~(\ref{eqn:thermaldist})
times a relativistic Breit-Wigner distribution over an interval $[m-\delta m_l,
m+\delta m_u]$, where $\delta m_l = \min [m-m_{\rm threshold},2\Gamma]$ and
$\delta m_u = 2\Gamma$. The minimum mass $m_{\rm threshold}$ is required to
open all decay modes\footnote{In fact, in analysis A (see below) the integration 
interval has been taken symmetric $[m-\delta m_l, m+\delta m_l]$}. The relativistic
Breit-Wigner distribution has been renormalised within the integration interval.
The non-vanishing width of resonances plays a major role especially at low energies 
(e.g. SIS); for instance, the $\Delta(1232)$ resonance creates pions more effectively 
than in the case of a vanishing width.

We have not used proper volume corrections in a Van der Waals type fashion which have 
been considered previously \cite{gorenstein}.

A major problem in Eq.~(\ref{eqn:sum}) is where to stop the summation over
hadronic states. Indeed, as mass increases, our knowledge of the hadronic
spectrum becomes less accurate; starting from $\approx 1.7$ GeV many states
are possibly missing, masses and widths are not well determined and so are the
branching ratios. For this reason, it is unavoidable that a cut-off on
hadronic states be introduced in Eq.~(\ref{eqn:sum}). If the calculations are 
sensitive to the value of this cut-off, then the reliability of results is 
questionable. We have performed all our calculations with two cut-offs, one at 1.8 GeV
(in the analysis algorithm A) and the other one at 2.4 GeV (in the analysis
algorithm B). The contribution of missing heavy resonances is expected to be
very important for temperatures $\geq 200$ MeV making thermal models inherently
unreliable above this temperature.

\section{Results}

As mentioned in the introduction, we have performed two analyses (A and B)
by using completely independent algorithms. In the analysis A all light-flavoured 
resonances up to 1.8 GeV have been included.
The production of neutral hadrons with a fraction $f$ of \ssb content has been
suppressed by a factor $(1-f)+f\gamma_s^2$. In the analysis B the mass cut-off has
been pushed to 2.4 GeV and neutral hadrons with a fraction $f$ of \ssb content
have been suppressed by a factor $\gamma_s^{2f}$. Both algorithms use masses, widths
and branching ratios of hadrons taken from the 1998 issue of Particle Data
Table \cite{PDG}. However, it must be noted that differences between the two analyses
exist in dealing with poorly known heavy resonance parameters, such as assumed
central values of mass and width, where the Particle Data Table itself gives only
a rough estimate. Moreover, the two analyses differ by the treatment of mass windows
within which the relativistic Breit-Wigner distribution is integrated.

The results of the  $\chi^2$ fits are shown in Tables 1 and 2 for both analyses
A and B. The agreement is indeed very good and confirms the reliability of the
results obtained. The $\chi^2$ minimisation in Au--Au collisions at AGS in analysis B
did not converge to a reliable minimum; however, the $\chi^2$ computed in analysis B 
by fixing the values of the parameters to the ones obtained in analysis A is
approximately as large as in analysis A itself, thus confirming the good agreement 
between the two calculations.

We have investigated in detail the lack of convergence of analysis B in Au--Au
collisions. The main reason of the fragility of the fit is the absence of measured
antibaryon yields, which are very effective in fixing the baryon-chemical potential,
in the main set of full phase space data. That shortage brings about a shallowness of 
$\chi^2$ minima in four dimensions, and, consequently, a difficult convergence in
both analyses.
Notwithstanding, in analysis A the absolute minimum turned out to be deep enough,
whereas in analysis B the convergence to a sufficiently nearby point was spoiled
and the minimum drifted to $T \approx 150$ MeV with a nearly flat descent from the
minimum found in A. This indicated a possible model dependence of the fit outcome.
In order to check our result in analysis A and make it robust we have repeated the 
fits in Au--Au collisions at AGS by using an additional measurement of 
${\rm\bar p}/{\rm p}$\cite{antipAu} ratio in the limited phase space region 
$1.0 < y < 2.2$ around midrapidity. The use of a ratio of particles measured by the 
same experiment under the same conditions reduces the involved systematic errors due
to slightly different centrality definitions (with respect to the other data set) 
and other possible sources. However, the actual ratio in full phase space might be 
different owing to different shapes of ${\rm\bar p}$ and p rapidity distribution and 
this effect has been taken into account by conservatively assigning a 20\% additional 
systematic error. The fit results are shown in Table~3; the two analyses are in very 
good agreement and, on top of that, the results for analysis A are in excellent
agreement with those in Table~1 obtained without using ${\rm\bar p}/{\rm p}$ ratio, 
thus confirming the good quality of the calculation. 
         
For each analysis an estimate of systematic errors on fit parameters have been
obtained by repeating the fit
\begin{itemize}
\item assuming vanishing widths for all resonances
\item varying the mass cut-off to 1.7 GeV in analysis A and to 1.8 GeV in
analysis B
\item for Au--Au at 1.7$A$ GeV/c, the volume $V$ has been varied to $V/2$
and to $2V$ (see discussion in Sect.~2)  
\end{itemize}
The differences between the  new fit parameters and the main parameters have been
conservatively taken as uncorrelated systematic errors to be added in quadrature
for each variation (see Table~1). The effect of errors on masses, widths and branching
ratios of inserted hadrons has been studied in analysis A according to the procedure
described in ref.~\cite{bgs} and found to be negligible.

Finally, the results of the two analyses have been averaged according to a method
suggested in ref.~\cite{schm}, well suited for strongly correlated measurements.
Firstly, a simple no-correlation weighted average has been calculated as
the central value of each parameter. Secondly, the error on it has been estimated by
conservatively assuming that the results A and B are fully correlated, i.e. with a 
covariance matrix:

\begin{equation}
 C = \left( \begin{array}{cc} \sigma_1 & \sigma_1 \sigma_2 \\
                              \sigma_1 \sigma_2 & \sigma_2 \\
            \end{array} \right)                      
\end {equation}
yielding an error:

\begin{equation}
\sigma^2 = {{\frac{1}{\sigma_1^2}+\frac{1}{\sigma_2^2}+
\frac{2}{\sigma_1 \sigma_2}}\over \left(\frac{1}{\sigma_1^2}+\frac{1}
{\sigma_2^2}\right)^2}
\end {equation}
The correlation between analyses A and B clearly arises from the use of the
same set of hadronic data and theoretical model.

\begin{figure}[tbh]
\centerline{
\psfig{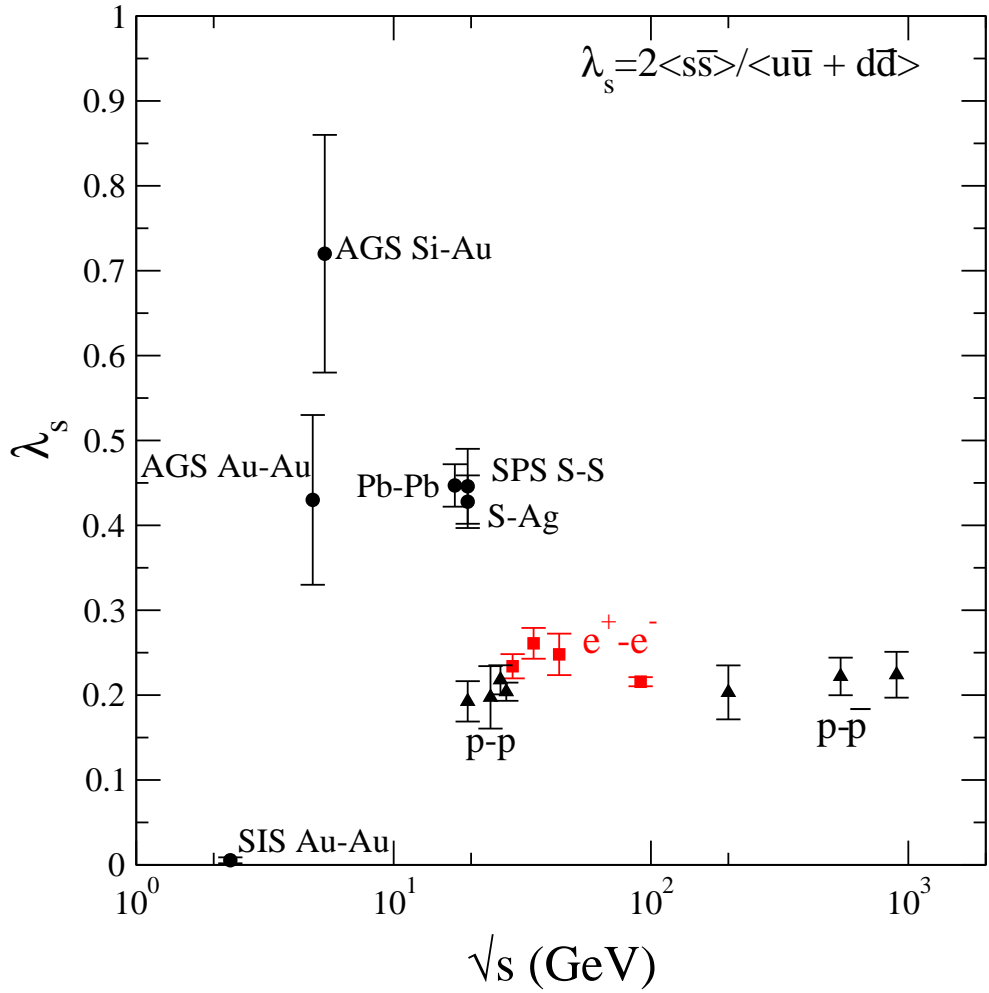}}
\vspace{1.5cm}
\caption{Ratio of strange quark pairs to created non-strange
quark pairs (Wroblewski factor) as a function of nucleon-nucleon
centre-of-mass energy. Values for S--S, S--Ag and \ee, pp, \ppb collisions
have been taken from ref.~\cite{bgs}.}
\label{fig:ls}
\end{figure}

In Table~1 we also list the values of the Wroblewski factor \ls \cite{wroblewski}
measuring the number of newly created {\em primary} valence \ssb pairs in
comparison to the newly created non-strange primary valence quark pairs

\begin{equation}
\lambda_{\rm s}=
 {2 \left<{\rm s\bar{s}}\right>\over \left< {\rm u\bar{u}}\right> +
\left< {\rm d\bar{d}}\right>}
\label{eqn:wrobl}
\end{equation}
along with fit and systematic errors. The \ssb and light quark pairs are computed 
on the basis of primary multiplicities of all hadron species, i.e. before particle 
decays take place. The behaviour of \ls as a function of collision type and 
centre-of-mass energy is shown in Fig.\ \ref{fig:ls} including elementary and S--S, 
S--Ag collisions. Values for S--S, S--Ag and \ee, pp, \ppb collisions have been 
taken from ref.~\cite{bgs}.

\section{Discussion and conclusions}

From the results obtained, it emerges that a statistical-thermal description of 
multiplicities in a wide range of heavy ion collisions is indeed possible to a 
satisfactory degree of accuracy, for beam momenta ranging from 1.7$A$ GeV/c to 
158$A$ GeV/c per nucleon. Furthermore, the fitted parameters show a remarkably 
smooth and consistent dependence as a function of centre-of-mass energy. The fit 
quality is generally good with the exception of Au--Au collisions at 1.7$A$ GeV/c 
where the large $\chi^2$ is due to an underestimation of one ratio $\eta/\pi^0$ 
(see Table~2).

\begin{figure}[tbh]
\centerline{
\psfig{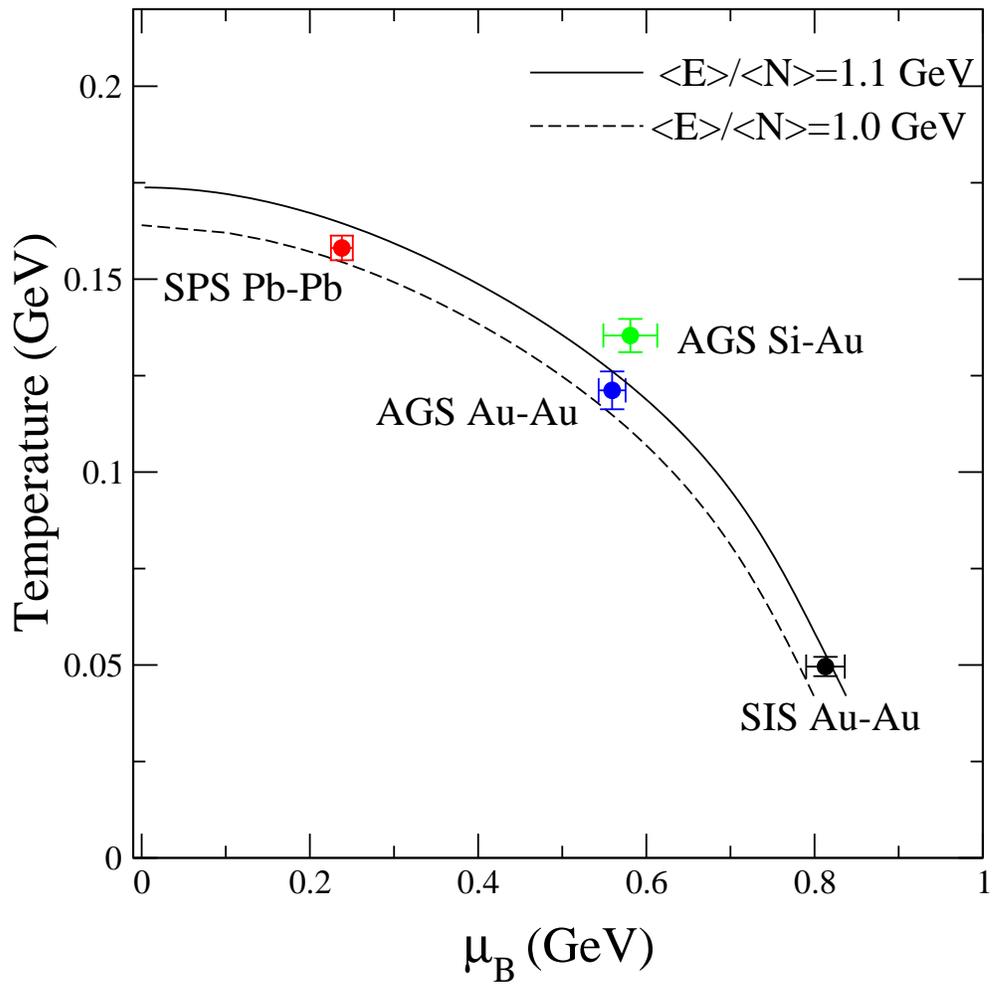}}
\vspace{1.5cm}
\caption{Fitted temperatures and baryon-chemical potentials plotted
along with curves of constant energy per hadron.}
\label{fig:eovern}
\end{figure}

The temperature varies considerably between the lowest and the highest beam energy,
namely, between 50 MeV at SIS and 160 MeV at SPS. Similarly, the baryon chemical
potential changes appreciably, decreasing from about 820 MeV at SIS to about 240
MeV at SPS. However, since the changes in temperature and chemical potential are
opposite, the resulting energy per particle shows little variation and remains
practically constant at about 1 GeV per particle; this is shown in Fig.\ \ref{fig:eovern}.

The supplementary \gs factor, measuring the deviation from a completely equilibrated 
hadron gas, is around 0.7 -- 0.8 at all energies where it has been considered a free 
fit parameter. At the presently found level of accuracy, a fully equilibrated hadron 
gas (i.e. \gs=1) cannot be ruled out in all examined collisions except in Pb--Pb, 
where \gs deviates from 1 by more than $4\sigma$.
This result does not agree with a recent similar analysis of Pb--Pb data \cite{heppe} 
imposing a full strangeness equilibrium. The main reason of this discrepancy is to 
be found in the different data set used; whilst in ref.~\cite{heppe} measurements in 
different limited rapidity intervals have been collected, we have used only particle 
yields extrapolated to full phase space. The temperature values that we have found 
essentially agree with previous analyses in Au--Au collisions \cite{cor} and Si--Au 
collisions \cite{BNL-thews,heppe,BNL-becattini} and estimates in Au--Au collision at
11.7 $A$ GeV/c \cite{stachel}.

The $T$ value in Pb--Pb is significantly affected by the multiplicity value of the
heaviest particles measured, namely $\phi$ and $\Xi$, as they are almost entirely
directly produced and provide a major lever arm on the slope of production
vs. mass function. A recent 40\% lowering of the $\Xi^-$ yield measured by NA49
\cite{barton} with respect to a previous measurement \cite{NA49-xi} results in a 
decrease of estimated temperature value from about 180 MeV to the actual 160 MeV. 
However, the removal of these two particles from the data set yields fitted parameter 
values which are in fair agreement with the main fit, as shown in Table~3. In 
particular, it is worth remarking that this exclusion does not bring significant
change to \gs whose outcome is very sensitive to particles with multiple strange
quark content and this confirms again the robustness of the main fit.   

In order to further investigate strangeness production in Pb-Pb we have performed
a consistency test between our fitted parameters, based on NA49 measurements, and
the multiplicities of multi-strange hadrons measured by the experiment WA97 in
central Pb--Pb collisions in a rapidity window $\pm 0.5$ around mid-rapidity
\cite{wa97}. By fixing $T$, \gs and $\mu_B$ to the averaged values in Table~1 and 
adjusting the volume (i.e. an overall normalisation), we obtain a $\chi^2/{\rm dof} 
= 28.9/6$. Calculated $\Lambda, \bar\Lambda$ multiplicities (see Table~4) do not 
include a residual feeding from $\Xi$ decays in the experiment, estimated to be 
$<5\%,<10\%$ respectively\cite{wa972}. The high value of the $\chi^2$ indicates
that the statistical-thermal analysis is not able to reproduce data in a limited
phase space region and in full phase space at the same time without resorting to
a more detailed dynamical model. In particular, the parameters determined by
the fit to NA49 data underestimate the yields of $\Xi$ and $\Omega$ baryons. 

\begin{figure}[tbh]
\centerline{
\psfig{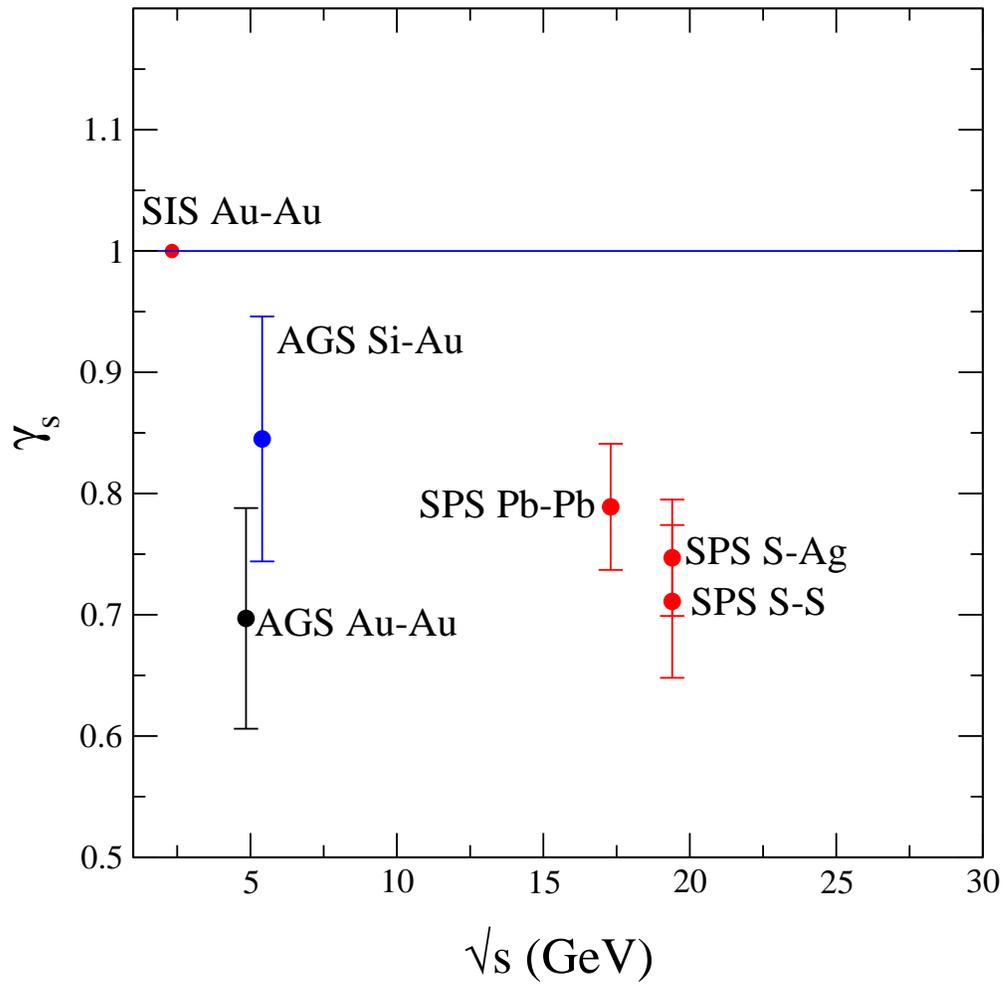}}
\vspace{1.5cm}
\caption{Strangeness suppression factor \gs as a function of nucleon-nucleon
centre-of-mass energy. Values for S--S, S--Ag and \ee, pp, \ppb collisions 
have been taken from ref.~\cite{bgs}}
\label{fig:gs}
\end{figure}

The parameter \gs as a function of centre-of-mass energy in heavy ion collision
(including S--S and S--Ag \cite{bgs}) is shown in Fig.\ \ref{fig:gs}.
Again, the values for S--S, S--Ag and \ee, pp, \ppb collisions have been taken 
from ref.~\cite{bgs}. As can be seen from the Fig.\ \ref{fig:gs} \gs  is
fairly constant, however, given the large error bars, it is quite difficult
to exclude different behaviours. Also the behaviour of \ls factor 
(see Fig.\ \ref{fig:ls}) as a function of energy (provided that there is little 
dependence on system size at fixed $\sqrt s$, as the approximate equality
of \ls in S--S and S--Ag confirms) is still unclear due to large experimental
uncertainties. The line shape is either compatible with a monotonically increasing
curve, saturating at \ls$\simeq 0.45$, or with a curve having a maximum around
Si--Au collisions, then decreasing and settling at an asymptotic $\simeq 0.45$
value or maybe decreasing further to the characteristic value of elementary
collisions.

Forthcoming lower energy Pb--Pb and high energy Au--Au data at RHIC should allow
to clarify the behaviour of strangeness production in heavy ion collision. In
order to easily compare our results with new measurements from RHIC experiments 
we also show in figure 4 the values of various particle-antiparticle
ratios as a function of ${\rm{\bar p}/p}$ ratio for different values of the 
temperature ($T$=160, 165 and 170 MeV) and a fixed value of the charge to baryon 
ratio of 0.401. RHIC results, however, will only be available for very limited 
kinematical region, while this kind of thermal model approach is largerly tied 
to full phase space ratios (see Introduction).

\section*{Acknowledgements}
We are very grateful to N. Carrer, U. Heinz, M. Morando, C. Ogilvie
for useful suggestions and discussions about the data. We especially
thank H. Oeschler for his help with the GSI SIS data and R. Stock for
his help with NA49 data.
\newpage

%

\newpage

\begin{table}[ht]
\begin{center}
\caption{Summary of fit results. Free fit parameters are quoted along with
resulting minimum $\chi^2$'s and \ls parameters.}
\vspace{0.5cm}
\begin{tabular}{|c|c|c|c|}
\hline
 & Analysis A        & Analysis B        &  Average\\
\hline
\multicolumn{4}{|c|}{Au--Au 1.7$A$ GeV} \\
\hline               
$T$ (MeV)       & 49.6$\pm$1.0$\pm$2.2 & 49.7$\pm$1.1$\pm$2.3  & 49.6$\pm$2.5 \\
$\mu_B$ (MeV)   & 810$\pm$15$\pm$12    & 818$\pm$15$\pm$12     & 813$\pm$23  \\
\gs             & 1 (fixed)            & 1 (fixed)             & 1 (fixed)   \\
$V$(fm$^3$)     & 1437 (fixed)         & 1437 (fixed)          & 1437 (fixed)  \\
\hline
$\chi^2$/dof    & 14.9/2               &  15.1/2               &  \\
\ls             & 0.0050$\pm$0.0034    &  0.0058$\pm$0.0036    & 0.0054$\pm$0.0035 \\
\hline
\multicolumn{4}{|c|}{Au--Au 11.6$A$ GeV} \\
\hline                
$T$ (MeV)             & 121.2$\pm$3.9$\pm$3.0     &    & 121.2$\pm$4.9  \\
$\mu_B$ (MeV)         & 559$\pm$15$\pm$9          &    & 559.4$\pm$16   \\
\gs                   & 0.697$\pm$0.080$\pm$0.043 &    & 0.697$\pm$0.091 \\
\vexp                 & 2.01$\pm$0.23$\pm$0.14    &    & 2.01$\pm$0.27  \\
\hline
$\chi^2$/dof          &  2.25/2           &     &                  \\  
\ls                   &  0.43$\pm$0.10    &     &  0.43$\pm$0.10   \\   
\hline
\multicolumn{4}{|c|}{Si--Au 14.6$A$ GeV} \\
\hline                
$T$ (MeV)             & 133.1$\pm$3.9$\pm$1.4   & 138.0$\pm$4.5$\pm$0.3   & 135.4$\pm$4.3    \\
$\mu_B$ (MeV)         & 592$\pm$34$\pm$13       & 573$\pm$30$\pm$0.8      & 581$\pm$32       \\
\gs                   &0.843$\pm$0.095$\pm$0.067&0.847$\pm$0.061$\pm$0.069& 0.845$\pm$0.101  \\
\vexp                 &0.526$\pm$0.090$\pm$0.081&0.545$\pm$0.079$\pm$0.12 & 0.534$\pm$0.130  \\
\hline
$\chi^2$/dof          & 14.3/4                  &  11.6/4                 & \\
\ls                   & 0.74$\pm$0.2\           &  0.72$\pm$0.12          & 0.72$\pm$0.14  \\
\hline               
\multicolumn{4}{|c|}{Pb--Pb 158$A$ GeV} \\
\hline 
$T$ (MeV)       & 159.5$\pm$2.5  $\pm$1.5  & 156.0$\pm$2.4$\pm$2.6    & 158.1$\pm$3.2 \\
$\mu_B$ (MeV)   & 238  $\pm$13   $\pm$3    & 239$\pm$12$\pm$5         & 238$\pm$13    \\
\gs             & 0.760$\pm$0.035$\pm$0.028& 0.862$\pm$0.036$\pm$0.061& 0.789$\pm$0.052   \\
\vexp           & 20.9 $\pm$1.5  $\pm$2.0  & 19.7$\pm$1.0$\pm$2.9     & 21.7$\pm$2.6      \\
\hline
$\chi^2$/dof    & 14.4/6                   & 22.6/6                   &  \\
\ls             & 0.444$\pm$0.026          & 0.450$\pm$0.024          & 0.447$\pm$0.025   \\
\hline
\end{tabular}
\end{center}
\end{table}
\newpage
\begin{small}
\begin{table}[t]
\begin{center}
\caption{Comparison between fitted and measured particle multiplicities
and ratios. In Au--Au collisions at AGS we also quote our prediction
 (including
weak decay products) along with a measurement in a limited kinematic range
($1.0 \le y \le 2.2$), whose error
 is only statistical, which sets a lower limit
for $\bar{\rm p}$ multiplicity.}
\vspace{0.5cm}
\begin{tabular}{|c|c|c|c|c|}
\hline
         & Reference   & Measurement & Analysis A & Analysis B \\
\hline 
\multicolumn{5}{|c|}{Au--Au 1.7$A$ GeV}\\
\hline                                       
$\pi^+$/p     & \cite{cor}   & 0.052$\pm$0.013  &    0.05306 & 0.05306  \\
K$^+/\pi^+$   & \cite{cor}   & 0.003$\pm$0.00075&    0.003040& 0.003030  \\
$\pi^-/\pi^+$ & \cite{cor}   & 2.05$\pm$0.51    &    2.0371  & 2.007     \\
$\eta/\pi^0$  & \cite{cor}   & 0.018$\pm$0.007  &    0.00109 & 0.000851  \\
\hline
\multicolumn{5}{|c|}{Au--Au 11.6$A$ GeV}\\
\hline  
Participants  &\cite{center-BNL} & 363$\pm$10      &   363.0    &      \\
K$^+$         &\cite{center-BNL} & 23.7$\pm$2.9    &   20.23    &      \\
K$^-$         &\cite{center-BNL} & 3.76$\pm$0.47   &   4.038    &      \\
$\pi^+$       &\cite{piAu}       & 133.7$\pm$9.9   &   133.3    &      \\
$\Lambda$     &\cite{LAu}        & 20.34$\pm$2.74  &   21.54    &      \\
p$/\pi^+$     &\cite{pAu}        & 1.234$\pm$0.126 &   1.295    &      \\
\hline
$\bar{\rm p}$ &\cite{antipAu}    &$>$0.0185$\pm$0.0018  & 0.0363  & \\
\hline
\multicolumn{5}{|c|}{Si--Au 14.6$A$ GeV}\\
\hline
Participants  & \cite{Si1}  & 115$\pm$10     & 99.41      &  94.33     \\
$\pi^+$       & \cite{Si1}  &  33$\pm$3      & 34.79      &  36.92     \\
$\pi^-/\pi^+$ & \cite{Si1}  & 1.09$\pm$0.13  & 1.296      &  1.196     \\
K$^+/\pi^+$   & \cite{Si2}  & 0.18$\pm$0.02  & 0.1564     &  0.1590    \\
K$^-/\pi^-$   & \cite{Si2}  &0.034$\pm$0.004 & 0.02715    &  0.02767   \\
$\bar{\rm p}$/K$^-$ & \cite{Si3}  &0.018$\pm$0.0034& 0.01672  & 0.01706  \\
$\bar{\Lambda}/\Lambda$& \cite{Si2}  &0.003$\pm$0.0015&0.00217 &0.00301   \\
$\phi$        & \cite{Si2}  &0.09$\pm$0.04   & 0.1725 &  0.1345 \\
\hline
\multicolumn{5}{|c|}{Pb--Pb 158$A$ GeV}\\                          
\hline
$(\pi^++\pi^-)/2.$   &\cite{sikler}   & 600$\pm$30   &  581.9 &  568.0    \\
K$^+$                &\cite{sikler}   & 95 $\pm$10   &  96.42 &  99.05    \\
K$^-$                &\cite{sikler}   & 50 $\pm$5    &  56.53 &  60.96    \\
K$^0_S$              &\cite{sikler}   & 60 $\pm$12   &  75.39 &  79.34    \\
p                    &\cite{sikler}   & 140$\pm$12   &  144.9 &  144.9    \\
$\bar{\rm p}$        &\cite{sikler}   & 10 $\pm$1.7  &  8.242 &  7.707    \\
$\phi$               &\cite{barton}   & 7.6$\pm$1.1  &  7.185 &  5.852    \\
$\Xi^-$              &\cite{barton}   & 4.42$\pm$0.31&  3.895 &  4.110    \\
$\overline{\Xi^-}$   &\cite{barton}   & 0.74$\pm$0.04&  0.766 &  0.765    \\
$\overline{\Lambda}/\Lambda$&\cite{sikler} & 0.2$\pm$0.04 & 0.1033 &0.098\\
\hline  
\end{tabular}
\end{center}
\end{table}
\end{small}
\begin{table}[th]
\begin{center}
\caption{Top: fit results for Au--Au collisions with the measured ratio ${\rm\bar p}/
{\rm p}$; its systematic error has been conservatively estimated to be 20\%.
Bottom: fit results for Pb--Pb collisions with the exclusion of $\phi$, $\Xi$'s and
both.}
\vspace{0.5cm}
\begin{tabular}{|c|c|c|c|}
\hline
\noalign{\smallskip}
\multicolumn{4}{c}{Au--Au 11.6$A$ GeV}\\
\noalign{\smallskip}
\hline    
                & Analysis A                & Analysis B               &  Measurement   \\
\hline           
$T$ (MeV)       & 121.2$\pm$4.6$\pm$1.7     & 130.6$\pm$5.5$\pm$3.9    &  \\
$\mu_B$ (MeV)   & 558$\pm$12$\pm$9          & 594$\pm$26$\pm$30        &      \\
\gs             & 0.701$\pm$0.068$\pm$0.072 & 0.883$\pm$0.124$\pm$0.207&      \\
\vexp           & 2.02$\pm$0.19$\pm$0.36    & 1.65$\pm$0.22$\pm$0.66   &   \\
\hline
$\chi^2$/dof    & 2.70/3                    & 1.06/3        &      \\ 
\ls             & 0.44$\pm$0.11             & 0.72$\pm$0.30 &      \\
Participants          & 363.0    &  364.1    &   363$\pm$10     \\
K$^+$                 & 20.23	 &  21.37    &   23.7$\pm$2.9  	   \\
K$^-$                 & 4.038	 &  3.950    &   3.76$\pm$0.47 	   \\
$\pi^+$               & 133.3	 &  130.5    &   133.7$\pm$9.9 	   \\
$\Lambda$             & 21.54	 &  21.40    &   20.34$\pm$2.74	\\
p$/\pi^+$             & 1.295	 &  1.237    &   1.234$\pm$0.126\\
$\bar{\rm p}/{\rm p}$ & 2.125\,10$^{-4}$& 2.516\,10$^{-4}$ &(2.50$\pm$0.25$\pm$0.50)10$^{-4}$\\
\hline        
\noalign{\smallskip}
\multicolumn{4}{c}{Pb--Pb 158$A$ GeV - Analysis A}\\
\noalign{\smallskip}
\hline 
               & without $\phi$     & without $\Xi$'s &  without $\phi$ and $\Xi$'s \\
\hline
$T$ (MeV)      & 159.9$\pm$2.6      &  168.7$\pm$7.1  &  165.6$\pm$6.5     \\
$\mu_B$ (MeV)  & 237  $\pm$13       &  222$\pm$15     &  218$\pm$9         \\
\gs            & 0.753$\pm$0.036    &  0.710$\pm$0.051&  0.664$\pm$0.064   \\
\vexp          & 22.8 $\pm$1.2      &  23.8$\pm$1.3   &  24.4$\pm$1.3      \\
\hline
$\chi^2$/dof   & 14.0/5    &   5.5/4                  &  4.5/3 \\
$\phi$         & 7.092     &   6.829                  &  6.053 \\
$\Xi^-$        & 3.849     &   3.616                  &  3.154 \\
$\bar\Xi^+$    & 0.7637    &   0.9749                 &  0.8028 \\
\hline
\noalign{\smallskip}
\multicolumn{4}{c}{Pb--Pb 158$A$ GeV - Analysis B}\\
\noalign{\smallskip}
\hline 
               & without $\phi$   & without $\Xi$'s   &  without $\phi$ and $\Xi$'s \\
\hline
$T$ (MeV)      & 158.2$\pm$2.7    &   167.1$\pm$6.3   &  158.5$\pm$5.2  \\
$\mu_B$ (MeV)  & 232$\pm$12       &   227$\pm$13      &  208$\pm$14     \\
\gs            & 0.806$\pm$0.040  &   0.862$\pm$0.043 &  0.658$\pm$0.067   \\
\vexp          & 20.9 $\pm$1.1    &   20.2$\pm$1.1    &  22.9$\pm$1.3  \\
\hline
$\chi^2$/dof   & 15.9/5        &   14.6/4                 &  6.3/3 \\
$\phi$         & 4.784         &   6.293                  &  2.328 \\
$\Xi^-$        & 3.867         &   4.397                  &  2.661 \\
$\bar\Xi^+$    & 0.7594        &   1.151                  &  0.5916 \\
\hline
\end{tabular}
\end{center}
\end{table}
\begin{table}[h]
\begin{center}
\caption{Comparison between predicted particle multiplicities in central Pb--Pb
collisions by using fitted $T$, $\mu_B$ and \gs quoted in the rightmost column
of Table~1 and those measured by WA97. The normalisation volume has been adjusted 
to minimise a $\chi^2$ which turned out to be 28.9.}
\vspace{0.5cm}
\begin{tabular}{|c|c|c|}
\hline
                    & Calculated     & Measured       \\
\hline                                                   
h$^-$               &  207.9         & 178$\pm$22     \\
K$^0_S$             &  23.66         & 21.9$\pm$2.4   \\
$\Lambda$           &  15.58         & 13.7$\pm$0.9   \\
$\bar\Lambda$       &  1.543         & 1.8$\pm$0.2    \\
$\Xi^-$             &  1.251         & 1.5$\pm$0.1    \\
$\bar\Xi^+$         &  0.2354        & 0.37$\pm$0.06  \\
$\Omega+\bar\Omega$ &  0.1662        & 0.41$\pm$0.08  \\
\hline  
\end{tabular}
\end{center}
\end{table}
\newpage

\newpage

\newpage

\newpage

\begin{figure}[tbh]
\centerline{
\psfig{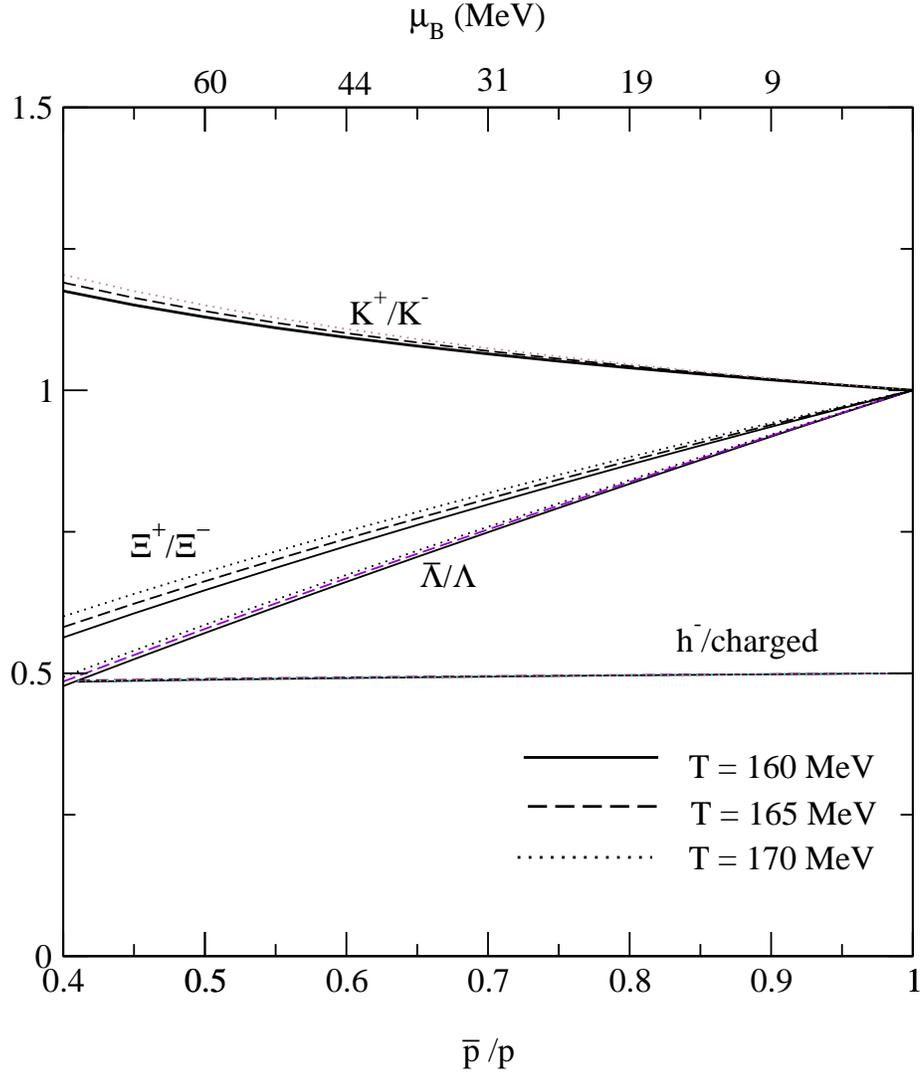}}
\vspace{2.5cm}
\caption{Particle-Antiparticle and negative to charged hadrons ratios as
a function of the ${\rm \bar{p}/p}$ ratio for different temperatures and a 
fixed ratio of charge over baryon number Q/B 0.401 and \gs=1. The 
${\rm \bar{p}/p}$ variation is governed by the variation of the baryon 
chemical potential. The dependence of these ratios on \gs, within the 
expected range, is found to be negligible.}
\label{fig:rhic}
\end{figure}

\begin{thebibliography}{999}
\bibitem{heinz} U. Heinz, Nucl. Phys. {\bf A661}, 140c (1999).
%
\bibitem{gs} J. Letessier, J. Rafelski, A. Tounsi, Phys. Rev. C{\bf 64},
406, (1994);
C. Slotta, J. Sollfrank, U. Heinz, Proc. of Strangeness in Hadronic matter,
J. Rafelski (Ed.), AIP Press, Woodbury 1995, p. 462.
%
\bibitem{oeschler} see e.g. H. Oeschler, Lecture Notes in Physics 516,
"Hadrons in dense matter and hadrosynthesis", Springer-Verlag (1999),
Eds. J. Cleymans, H.B. Geyer, F.G. Scholtz.
%
\bibitem{ogilvie} see e.g. C. Ogilvie, Nucl. Phys. {\bf A638}, 57c (1997).
%
\bibitem{stock} see e.g. R. Stock, Nucl. Phys. {\bf A661}, 282c (1999).
%
\bibitem{prc} J. Cleymans and K. Redlich, Phys. Rev. C{\bf 60}, 054908 (1999).
%
\bibitem{bgs} F. Becattini, M. Ga\'zdzicki and J. Sollfrank, Eur. Phys. J. 
C{\bf 5}, 143 (1998).
%
\bibitem{stockplb} R. Stock, Phys. Lett. B{\bf 456}, 277 (1999).
%
\bibitem{BNL-PBM} P. Braun-Munzinger, J. Stachel, J.P. Wessels and N. Xu,
Phys. Lett., B{\bf 344}, 43 (1995).
%
\bibitem{BNL-thews} J. Cleymans, D. Elliott, R.L. Thews and H. Satz,
Z. Phys. C{\bf 74}, 319 (1997).
%
\bibitem{abbottnew} T. Abbott et al., Phys. Rev. C{\bf 50}, 1024 (1994).
%
\bibitem{Si2} Y. Akiba et al., E-802 collaboration, Nucl. Phys. {\bf A590}, 
179c (1995).
%
\bibitem{heppe} P. Braun-Munzinger, I. Heppe and J. Stachel, Phys. Lett., 
B{\bf 465}, 15 (1999).
%
\bibitem{letessier} J. Letessier and J. Rafelski, Nucl. Phys. {\bf A661}, 
97c (1999).
%
\bibitem{Si3} T. Abbott et al., E-802 collaboration, Nucl. Phys. {\bf A525}, 
455c (1994).
%
\bibitem{piAu} L.Ahle et al., E-802 Collaboration, Phys. Rev. C{\bf 59},
2173 (1999).
%
\bibitem{LAu} S. Ahmad et al., Phys. Lett. B{\bf 382}, 35 (1996).
%
\bibitem{pAu} L.Ahle et al., E-802 Collaboration, Phys. Rev. C{\bf60}, 06
49001 (1999).
%
\bibitem{prl} J. Cleymans and K. Redlich, Phys. Rev. Lett. {\bf 81}, 5284
 (1998).
%
\bibitem{beca} F. Becattini , Lecture Notes in Physics 516, "Hadrons in dense 
matter and hadrosynthesis", Springer-Verlag (1999), Eds. J. Cleymans, H.B. Geyer, 
F. G. Scholtz.
%
\bibitem{PDG} Particle Data Group, Eur. Phys. J. C{\bf 3}, 1 (1998).
%
\bibitem{sikler} F. Sikler (NA49 Collaboration), Nucl. Phys. A{\bf 661}, 45c 
(1999).
%
\bibitem{antipAu} L.Ahle et al., E-802 Collaboration, Phys. Rev. Lett. {\bf 81}, 
2650 (1998).
%
\bibitem{canstran} J. Cleymans, D. Elliott, A. Keranen, E. Suhonen, Phys. Rev. 
C{\bf 57}, 3319 (1998).
%
\bibitem{gorenstein} G.D. Yen and M.I. Gorenstein, Phys. Rev. C{\bf 59},
2788 (1999).
%
\bibitem{cor} J. Cleymans, H. Oeschler and K. Redlich, Phys. Rev. C{\bf 59}, 
1663 (1999) and references therein.
%
\bibitem{center-BNL} L.Ahle et al., E-802 Collaboration, Phys. Rev. C{\bf 60}, 
044904 (1999).
%
\bibitem{Si1} T. Abbott et al., E-802 collaboration, Phys. Rev. C{\bf 50}, 1024
(1994).
%
\bibitem{barton} R.A. Barton, NA49 Collaboration, talk given at the 
``Strangeness 2000'' conference, Berkeley, (July 1999), to be published in the 
proceedings of the conference.
%
\bibitem{schm} M. Schmelling, Phys. Scripta {\bf 51}, 676 (1995).
%
\bibitem{wroblewski} A. Wroblewski, Acta Physica Polonica, B{\bf 16}, 379 (1985).
%
\bibitem{BNL-becattini} F. Becattini, J. Phys. G{\bf 25}, 287 (1999).
%
\bibitem{stachel} J. Stachel, Nucl. Phys. {\bf A610}, 509c (1996).
%
\bibitem{NA49-xi} H. Appelshauser et al., NA49 Collaboration, Phys. Lett. 
B{\bf 444},523  (1998).
%
\bibitem{wa97} F. Antinori, Nucl. Phys. {\bf A661}, 130c (1999);
D. Elia, Nucl. Phys. {\bf A661}, 649c (1999); the measured multiplicities have 
been taken from the web page of the WA97 experiment http://www.cern.ch/WA97/QM99table/TableQM
99.html (M. Morando, private communication).
%
\bibitem{wa972} F. Antinori et al., Eur. Phys. J. C{\bf 14}, 633 (2000).
%
\end{thebibliography}
\end{document}